# TRAJECTORY-WAVE APPROACH TO ELECTRON DYNAMICS IN HYDROGEN ATOM

N.T.Valishin[1,2], F.T.Valishin[2] and S.A.Moiseev[3]

1) Kazan State Technical University (KAI), Kazan, Russia

2) Philosophy Methodology Center of Dynamism of Tatarstan Academy of Sciences, Kazan, Russia

3) Kazan Physical-technical Institute of Russian Academy of Sciences Sibirsky Trakt 10/7,

Kazan, 420029, Russia: E-mail: samoi@yandex.ru

**Abstract.**

In this work we propose a new approach to the explanation of the nature of electron based on the corpuscular-wave monism using the further development of the optical-mechanical analogy to describe the physical reality. In this theory the motion of an electron is considered to occur along a trajectory the presence of which is a reflection of the existence of a particle, as well as it is assumed that any motion is defined by a wave V(x,t). It is assumed that there is an explicit relationship between the trajectory and wave equations of the electron, which are established on the basis of the local variational principle. In this approach, an electron wave propagating in free space takes along the electron trajectory. We used this theory to describe the electron motion in a hydrogen-like atom and found its stationary states. The energies of these states coincide with the known quantum mechanics solutions for the stationary energies of the hydrogen-like atom, however, in our approach the spatial trajectories of the electron have the form of the surfaces, which are formed in the region of nodes of the standing electron wave. These surfaces have the form of spheres for the spherical symmetrical electron states and the radii of these spheres coincide with the radii of the Bohr orbits of these states. Thus, in this approach the trajectory and wave measurements of the electron get a consistent spatial description that is inherent to the picture of the corpuscular-wave monism. We discuss the considerable correspondence of the proposed theory with the quantum mechanics results describing the stationary and non-stationary motions of the electron in the atom and their difference. We believe that measuring of the spatial configuration of the observed electron trajectory surfaces in an atom could be a deep examination of the standard quantum theory.

**PACS-numbers**: 03.65.-w; 03.65.Ge; 67.63.Gh; 67.80.fh.

## 1. Introduction

Modern quantum theory was conceived in the works of M. Planck [1]. It resulted from a synthesis of the corpuscular and wave approaches to interpreting the spectrum of an absolutely black body that led to the discovery of the universal Planck constant h. In the following works, A. Einstein [2], N. Bohr [3] and Louis de Broglie [4] continued the study of the corpuscular and wave properties in the behavior of the electromagnetic field and electron leading them to the concept of a photon, quantization of the atom energy levels and physical de Broglie waves. E. Schrödinger [5] was most successful in the mathematical generalization of the Louis de Broglie wave theory having formulated an equation for the electron wave function in an arbitrary external field which was named after him. The mathematical fundamentals of quantum mechanics were formulated earlier in the terms of matrix mechanics in the works of W. Heisenberg [6], M. Born, W. Heisenberg, and P. Jordan [7] resulting from a need to construct a theory of the electron on basis of the use of such quantities observed in experiment as frequencies of the atom radiation and transition matrix elements between quantum states of the atom. Soon E. Schrödinger [8] and C. Eckart [9] have established the complete mathematical equivalence of both approaches using different quantities observed in experiment.

As creators of the quantum theory acknowledged, the wave approach allowed one to reach a deeper physical understanding of the nature of the phenomena of microcosm, however, the wave function of the electron had the non-classical nature and according to the proposal of M. Born, it was used to describe the amplitude of the probability of finding the electron in space [10]. This statistical interpretation allowed one to mathematically consistently combine matrix mechanics with the wave one, explaining corpuscular and wave properties of the electron observed in experiment. Thus, matrix and wave mechanics were developed into modern quantum theory that up to now described successfully a large number of experimental facts. The quantum theory of light was built by P. Dirac somewhat later, following the projecting of the main statements of the quantum theory on the systems with a continuous number of the degrees of freedom [11] that led to discovery of the physical vacuum as well.

The probabilistic interpretation of the wave function was an unexpected result even for creators of the quantum theory themselves. However, this interpretation followed indivertibly from the absence of the classical trajectories of the electron in the wave equation of E. Schrödinger and in the following equations of the quantum theory built on similar principles. In turn, the inconsistency of the trajectory and wave descriptions (corpuscular-wave dualism) was considered as one of the main postulates of the theory philosophically motivated by the Heisenberg uncertainty relation [12] and manifested by the appearance of quantum operators satisfying the commutation relations for the conjugated physical quantities in the theory [5,7, 8]. According to the spirit of the mathematical apparatus of the wave theory, the quantum superposition principle was incorporated in quantum mechanics for the wave function of the electron (photon, etc.), which states the possibility of the "simultaneous" coexistence of different classical trajectories of the motion of the electron from one point in space to another. This principle was the basis of the Feynman formulation of the quantum theory [13] and became one of its key statements radically contradicting the classical picture of the physical reality.

The creation of quantum mechanics as an especially non-classical theory was anyway a result of the comprehension of a complex and inconsistent picture of the manifestation of the corpuscular and wave properties in the behavior of the electron (photon, etc.), obviously not fitting the Newton—Maxwell picture of the behavior of microparticles and light. The longing to reach a more complete understanding of the corpuscular and wave properties in the studied systems, however, exists at present as well, although mainly following spirit of the N. Bohr's complimentarily principle. This approach is successfully manifested in development of the quantum detection theory [14]; Young's experiments on quantum interference on complex quantum systems, for instance, molecules containing a large number of atoms [15]; study of the quantum systems in quantum mixed states unusual for classics, for instance, states of the Schrödinger cat type [16-18], combining in one the microscopic and macroscopic subsystems; studies of optical quantum interference of photons in a medium at the most general physical conditions [19]; upon the development of new quantum ways of the information transmission [20] and processing [21]. Probably, the brightest results were achieved by the quantum theory of measurements in the description of the experiments on continuous observation of single quantum objects (atoms and molecules in different traps, tunneling microscopes etc.), which to a considerable extent became possible due to the achievements of modern quantum optics [22-26], where the experiments with single photons and atoms which recently seemed to be absolutely impossible are available.

Recent experimental achievements in the study of the behavior of separate microscopic systems, in turn, revive the stable interest in testing the main statements of the quantum theory and stimulate a deeper re-comprehension of its physical fundamentals and the role of information in the theoretical description of the behavior of microparticles [26, 27]. The continuing attempts to understand the paradoxical manifestations of the corpuscular-wave dualism in the motion of the electron (and other microparticles) also stimulate the creation of new theories anyway developing the Louis de Broglie's ideas of the wave-pilot [28-30], in spite of the impossibility of its naive implementation according to the modern achievements of the quantum theory [31]. In this work we propose a new approach to solving this problem based on the corpuscular-wave monism in the explanation of the nature of electron. Namely, the theory developed below uses the description of the physical reality, in which it is taken into account that there are electron trajectories serving as a reflection of the existence of the particle. At the same time, it is also assumed that the motion the electron is determined by the physical wave $V(x,t)$. It should be noted that unlike the positivist approach [6, 7] used to build quantum mechanics and based on the description of reality only by means of quantities observed in experiment (transition dipole moments, frequencies of radiated photon, etc. manifesting the way of the existence of electron), we use the notion "process—state" which is introduced to describe the essence and way of the existence of electron. This notion is initially formulated on the basis of the ontology from the strategy of dynamism [32] in which the motion (wave process) is the essence of reality and the trajectory (state) is the way of the existence of reality. We show below how the introduction of the notion "process—state" allows one to describe the wave and corpuscular measurements in the behavior of electron within a unique spatial picture.

The proposed theory is developed using the generalization of the optical-mechanical analogy to the description of the trajectory and wave behavior of electron. At first, the main statements of the corpuscular-wave monism are formulated and their physical meaning is clarified on the basis of using the local variational (LV) principle [33]. Then we use this theory to describe the electron in free space and the stationary Coulomb field of the hydrogen-like atom, one of the test objects of the quantum theory. In conclusion, the obtained results are discussed and the discovered physical picture of the behavior of the electron is clarified on their basis. Finally, we outline briefly the opening possibilities in the description of the new manifestations of the corpuscular and wave nature of microobjects.

## 2. Local variational principle and the V(x,t) function method

Let us define the contents of the local variational (LV) principle. Let us specify the trajectory motion of an object by a system of the differential equations of classical physics:

$$\frac{d}{dt}x = f(x) \tag{1}$$

where the vector of the phase coordinates of the particle $x(t) = (x_1, x_2, ..., x_n)^T$ is given in the n-dimensional Euclidean space ($x \in R^n$), t is time.

In addition to the system of equations (1) also we introduce a wave function V(x,t). The velocity of its change for the studied system (1) is determined by the expression $\frac{d}{dt}V = \frac{\partial}{\partial t}V + \frac{\partial}{\partial x}V^T f$. We consider an isochronous variation of the velocity of the change of the wave function $\delta\left(\frac{d}{dt}V\right) = \frac{\partial}{\partial t}\delta V + \frac{\partial}{\partial x}\delta V^T f + \frac{\partial}{\partial x}V^T \delta f$, (where $\delta V = \frac{\partial}{\partial x}V^T \delta x$, $\delta f = \frac{\partial}{\partial x}f\delta x$). We assume that upon the variation of the speed of the change of the wave function $\delta\left(\frac{d}{dt}V\right)$ the object moves from a certain initial state to the state differing by a new spatial coordinate $x + \delta x$. We call such a transition the wave transition of the object in which the quantity $\delta V$ designates a possible wave transition from the initial state to the new state while $\delta x$ determines the trajectory variations. Upon the implementation of the wave transition, the spatial variation gets the form of the displacement $\delta x = dx = \dot{x} dt$ implemented in space.

Let us formulate the LV principle: *Of all possible transitions to the new state, there occurs the transition with the velocity of the change of the wave function V(x,t) at each instant having the stationary value*

$$\delta \left( \tfrac{d}{dt} V \right) = 0 \, . \tag{2}$$

By assuming that (2) holds, we also assume that the wave function satisfies the additional condition on the complete variation of the velocity of the change of the wave function V(x,t):

$$\widetilde{\Delta} \left( \tfrac{d}{dt} V \right) = 0 \, , \tag{3}$$

where $\widetilde{\Delta}(.) = \delta(.) + \tfrac{d}{dt}(.)\Delta t$.

Having the classical equations (1) and conditions (2), (3), we find the wave equation for V(x,t), taking into consideration the implementation of the wave transition ($\delta x = dx = \dot{x} dt$) in (2) and (3):

$$\widetilde{\Delta}\left(\frac{dV}{dt}\right) = \left\{ \frac{\partial^2 V}{\partial t^2} + 3 \frac{\partial^2 V}{\partial t \partial x}^T f + 2 f^T W f + 2 \frac{\partial V}{\partial x}^T \frac{df}{dt} \right\} dt$$

$$= 3\delta\left(\frac{dV}{dt}\right) + \left( \frac{\partial^2 V}{\partial t^2} - f^T W f - \frac{\partial V}{\partial x}^T \frac{df}{dt} \right) dt \to$$

$$\frac{\partial^2 V}{\partial t^2} - f^T W f - \frac{\partial V}{\partial x}^T \frac{df}{dt} = 0 \, , \tag{4}$$

where V(x,t) is a doubly continuously differentiable, finite, single-valued function, $W = [\partial^2_{x_i x_j} V(x, t)]$ - is a matrix of functions. Equation (4), according to [33], is the necessary and sufficient condition of the feasibility of (3). Let us show that the equality

$$\frac{\partial V}{\partial x}^T \frac{d}{dt} \dot{x} = 0 \tag{4a}$$

holds. According to the method of the V-function, the motion the particle occurs in such a manner that at each instant the velocity of the particle is directed along the gradient of the wave function i.e. $\tfrac{\partial}{\partial x} V^T \dot{x} = \left| \tfrac{\partial}{\partial x} V \right| |\dot{x}|$. Thus we obtain $\partial V / \partial x = k_2(x) \dot{x}$. We assume below that

the field of velocities in the three-dimensional space coincides with the corresponding gradient field that exists at $k_2(x) = k_2$ and, respectively, we obtain the equality

$$\partial V / \partial x = k_2 \dot{x}. \tag{4.b}$$

In the case when the wave transition is implemented, the relation (2) becomes

$$\frac{d}{dt}\left(\frac{\partial V}{\partial x}^T \delta x\right) = \frac{d}{dt}\left(\frac{\partial V}{\partial x}^T \dot{x} dt\right) = 0 \Rightarrow \frac{\partial V}{\partial x}^T \dot{x} = const. \tag{4c}$$

Then, allowing for (4b) and (4c), the equality (4a) holds, i.e.

$$\frac{\partial V}{\partial x}^T \frac{d}{dt}\dot{x} = k_2 \dot{x}^T \frac{d}{dt}\dot{x} = \frac{k_2}{2}\frac{d}{dt}(\dot{x}^T \dot{x}) = \frac{1}{2}\frac{d}{dt}(\frac{\partial V}{\partial x}^T \dot{x}) = 0.$$

As a result, equation (4) allowing for (4a) becomes

$$\partial^2_{tt} V - \dot{x}^T W \dot{x} = 0 \quad . \tag{5}$$

Equations (1) and (5) describe the trajectory and wave motion the studied particle. To find the solution of this system of equations it is necessary to know the boundary conditions. Note that we use the properties of the wave on the trajectory of the particle and on the boundary of the studied region of space determined by the method of the V(x,t) function as boundary conditions for (1) and (5). For the comparison, note that in classical physics the description of the dynamics of the particle is limited to equation (1) where the initial coordinate and the velocity of the particle in a certain fixed instant are designated. In quantum mechanics, in turn, only the Schrödinger wave equation is used, where the definite requirements to the wave function in the studied region of space determined in accordance with the general statements of the quantum theory are used as boundary conditions. The proposed approach to the description of the behavior of the particle contains a system of the trajectory equation (1) and the wave equation (4) or (5). Below we find the boundary conditions for the wave V(x,t) close to the trajectory of the particle.

Taking into account the implementation of the wave transition in (2), we obtain

**1st condition**

for the properties of the wave on the trajectory of the particle

$$\frac{\partial}{\partial x} V^T \dot{x} = const. \tag{6}$$

Taking into account that the motion the particle occurs on the trajectory along the direction of the gradient of the wave function as follows from the theorem about the wave and trajectory [32], we find that on the trajectory the following equality $\frac{\partial}{\partial x} V^T \dot{x} = \left|\frac{\partial}{\partial x} V\right| |\dot{x}|$ holds as well. Thus we obtain the boundary condition for the wave in the initial (boundary) point ($x = x_M$) of the trajectory of the particle $\partial V / \partial x \big|_{x=x_M} = k_2 \dot{x}\big|_{x=x_M}$.

**2nd condition**

Bearing in mind the implementation of the wave transition and condition (6), for the full variation (2), in turn, we obtain the equality $\frac{d}{dt}\left(\frac{\partial}{\partial t} V\right) = 0$. Using it we find, respectively, the 2nd condition for behavior of the wave on the trajectory of the particle $\frac{\partial}{\partial t} V = k_1$, where $k_{1,2}$ are constants.

**3rd condition**

on the properties of the wave V(x,t) follows from the condition of connectivity of the wave and trajectory. In this case the wave amplitude V(x,t) is zero in the point of the location of the particle (with coordinate $x = x_M$ at the instant $t$) $V(x = x_M, t) = 0$.

Thus, by summing, we write the general system of equations of the trajectory-wave motion of the particle according to the method of the V-function.

$$\tfrac{d}{dt} x = f(x),  \qquad (7.1)$$

$$\tfrac{\partial^2}{\partial t^2} V - f^T W f = 0,  \qquad (7.2)$$

complemented by the local relations for the wave and trajectory of the particle

$$\partial V / \partial x \big|_{x=x_M} = k_2 \dot{x}\big|_{x=x_M},  \qquad (7.3)$$

$$\tfrac{\partial}{\partial t} V = k_1,  \qquad (7.4)$$

$$V(x = x_M, t) = 0.  \qquad (7.5).$$

Note that the condition (7.3) is a particular case of (4.b) and is introduced as a boundary condition in order to use information available in each problem about the velocity of the particle in any part (or on the boundary) of space ($x \subseteq x_M$). Below the conditions (7.3) and (7.4) relating the behavior of the wave and particle close to its trajectory are specified and used to describe the behavior of the wave and particle for the free and limited in space motion of the electron. In turn, note that (7.5) is a condition of the existence of the trajectory of the particle supplementary to (7.3).

We consider the functioning of the system of equations (7.1)-(7.5) in the case of the uniform rectilinear motion of the particle, which, in turn allows us to define the physical meaning of the constants $k_1$ and $k_2$ entering this system of equations.

### 3. Uniform motion with constant velocity.

For the rectilinear motion of the electron with the constant velocity $\dot{x} = \upsilon$ equation (7.2) allowing for (7.1) has the following form:

$$\frac{\partial^2}{\partial t^2} V - \upsilon^2 \frac{\partial^2}{\partial x^2} V = 0 .$$

Note that the velocity $\upsilon$ in the wave equation coincides with the velocity of the particle motion. The possible solutions of the wave equation satisfying the marginal condition (7.3)-(7.5) have the following form:

$$V_1(x,t) = A \sin[\omega(x/\upsilon - t - T_o)] , \tag{8.1}$$

where $T_o = x_o/\upsilon - t_o$, as well as

$$V_2(x,t) = A \cos\{\omega(x/\upsilon - t)\} , \tag{8.2}$$

where the condition $\omega(x_o/\upsilon - t_o) = \pi/2 + \pi m$ holds, A is the wave amplitude the physical meaning of which is established by taking into account the relation (7.3). The latter takes into account that the particle motion occurs along the direction of the gradient of the wave function. Considering below only the solutions of (8.2), we obtain

$$\tfrac{\partial}{\partial x} V(x,t) = \tfrac{\partial}{\partial x} V_2(x,t) = (\omega/\upsilon) A \sin[\omega(x/\upsilon - t)] = k_2 \upsilon , \tag{9.1}$$

where the condition (7.4) is:

$$\tfrac{\partial}{\partial t} V(x,t) = -\omega A \sin[\omega(x/\upsilon - t)] = k_1 . \tag{9.2}$$

Since the right-hand parts of the last two equalities are real numbers, we obtain

$$\omega(x/\upsilon - t) = \text{Const} + \pi m . \tag{10}$$

Thus, the condition $x = x(t)$ always holds on the trajectory of the particle $V(x = x(t), t) = 0$, i.e., the particle as a point moves together with the wave on the trajectory experiencing no smearing in time. At the same time the wave amplitude is always zero in the particle location. Below, following the symmetry and simplicity considerations, we treat only the solution with $Const = \pi/2$. In this case, we find following relations for the wave amplitude

$$|A|\omega = k_2 \upsilon^2, \tag{11}$$

$$|A|\omega = k_1. \tag{12}$$

When the velocity of the motion is constant $\dot{x} = \upsilon$, according to (10) we find also the equation for the possible trajectory of the particle $x_n = \upsilon[t + (\pi/\omega)(n + 1/2)]$. In turn, for the chosen trajectory $t - x/\upsilon = C$ it follows from (10) that the carrier wave frequency is expressed via a certain minimum wave frequency analogously to the rule of quantization of the oscillator energy

$$\omega = \omega_n = \omega_o(n + 1/2), \tag{13}$$

where $\omega_o = \pi/|C|$ ($n = 1, 2, ...,$).

By assuming in (12) that $k_2 = m_e$, ($m_e$ is the electron mass), the A constant obtained the dimensionality of the action $[\kappa \varepsilon][M/c][M]$. Therefore we assume $A = h/2\pi = \hbar$, where $h$ is the Planck constant. From (11)-(12) we have $k_1 = m_e \upsilon^2 = 2E$, where $E$ is the electron energy. Thus, we obtain the following relations between the wave ($\upsilon, \lambda, \omega, A$) and trajectory ($\dot{x}, m, E$) properties of the particle motion

$$\upsilon = \dot{x}, \quad \omega = \frac{m_e \dot{x}^2}{\hbar} = \frac{2E}{\hbar}, \tag{14.1}$$

$$\lambda = \frac{h}{m_e \dot{x}}, \quad A = \hbar. \tag{14.2}$$

The relations (13) and (14.1) indicate the character of quantization of the frequency of the wave vibrations and the particle energy upon its uniform motion with the constant velocity $\upsilon$. According to (14.1), the energy is transferred by the particle. In turn, the particle momentum determines the wavelength (14.2), according to the known Louis de Broglie relation. As to the physical meaning, the wave $V(x, t)$ characterizes the properties of the action revealed in the electron motion. Thus, the wave by its node is connected with the particle location and thus leads it, at the same time the particle (trajectory) generates the wave propagating with it.

Note that the relations (14.1) and (14.2), inherent to the electron motion in free space, are used in the problem about the electron in the Coulomb field of the hydrogen-like atom considered below. The analysis of this problem opens new aspects in the manifestation of the relationship between the trajectory and wave motions of the electron in finite space, leading to the unique spatial picture of the electron trajectories and waves in the atom with the spectrum of the electron energy being quantized in accordance with the known properties of the hydrogen atom.

### 4. Electron in the Coulomb field

We consider the electron motion in the potential field. In this case, the trajectory equations of the object allow the existence of the integral of motion

$$\tfrac{1}{2} m_e (\dot{x}_1^2 + \dot{x}_2^2 + \dot{x}_3^2) + U(\vec{x}) = E , \qquad (15)$$

where $U(x)$ and $E$ are the potential and full energy of the electron.

In this case, equation (7.2) has the following form

$$\frac{\partial^2 V}{\partial t^2} - \sum_{n,m=1}^{3} \frac{\partial^2 V}{\partial x_n \partial x_m} \dot{x}_n \dot{x}_m = 0 . \qquad (16)$$

Using $\partial V / \partial x = k_2 \dot{x}$ when the condition of the coupling of the wave and trajectory $\dot{x}_j = \lambda_i \partial \dot{x}_i / \partial x_j$ $(i, j = \overline{1,3})$ holds [34], it is possible to establish that the equality

$$\frac{\partial^2 V}{\partial x_1^2}\left(\dot{x}_2^2 + \dot{x}_3^2\right) + \frac{\partial^2 V}{\partial x_2^2}\left(\dot{x}_1^2 + \dot{x}_3^2\right) + \frac{\partial^2 V}{\partial x_3^2}\left(\dot{x}_1^2 + \dot{x}_2^2\right) -$$

$$- 2 \frac{\partial^2 V}{\partial x_1 \partial x_2} \dot{x}_1 \dot{x}_2 - 2 \frac{\partial^2 V}{\partial x_1 \partial x_3} \dot{x}_1 \dot{x}_3 - 2 \frac{\partial^2 V}{\partial x_2 \partial x_3} \dot{x}_2 \dot{x}_3 = 0 \quad , \qquad (17)$$

holds.

Allowing for (17), equation (16) obtains the form of the three-dimensional wave equation

$$\tfrac{\partial^2}{\partial t^2} V - v^2 \Delta V = 0 , \qquad (18)$$

where the square of velocity is determined by the relation $v^2 = \sum_{i=1}^{3} \dot{x}_i^2 = 2(E - U(x))/m_e$ according to (15), $\Delta = \sum_{i=1}^{3} \frac{\partial^2}{\partial x_i^2}$ – is the Laplace operator. In this case, the wave equation (18) has the following form

$$\frac{\partial^2}{\partial t^2} V - \frac{2}{m}(E - U(x))\Delta V = 0. \tag{19}$$

Note again that instead of the solution of the equations of motion (1) with the given initial conditions on the velocity and coordinate of the particle, as in classical physics and the planetary Bohr model, we have the wave equation (19) (differing from the known Schrödinger equation), which should satisfy the boundary conditions on the wave properties. Equation (19) deserves particular attention due to its exclusively fundamental value. As we see below, its solutions give directly information about both the electron wave and trajectory.

By applying the variable division method $V = X(x)T(t)$ to solving equation (19), we obtain:

$$\frac{\frac{d^2}{dt^2} T(t)}{T(t)} = \frac{2(E - U(x))\Delta X(x)}{m_e X(x)} = -\omega^2. \tag{20}$$

From (20) we find the stationary equation

$$\frac{2(E - U(x))}{m_e}\Delta X + \omega^2 X = 0. \tag{21}$$

For the Coulomb field of the hydrogen-like atom $U(r) = -Ze^2/r$, equation (21) has the following form

$$\left(-\beta_0^2 + \frac{\alpha}{r}\right)\Delta X + \omega^2 X = 0, \tag{22}$$

where $\beta_0^2 = -2E/m$, $\alpha = 2Ze^2/m$. By transferring to the spherical coordinate system and keeping only the spherically symmetrical solutions when $X(\vec{r}) = R(r)$ (leaving the more general case of the arbitrary orbital motion for the further analysis), from (22) we obtain

$$\left(-\beta_0^2 + \frac{\alpha}{r}\right)\frac{1}{r^2}\frac{d}{dr}\left(r^2 \frac{dR}{dr}\right) + \omega^2 R = 0, \tag{23}$$

which after the change of variables $R = u/r$ has the following form

$$\frac{d^2u}{dr^2} + \left(\frac{k_0^2 \alpha}{\alpha - \beta_0^2 r} - k_0^2\right)u = 0, \tag{24}$$

where new constants $k_0^2 = \omega^2/\beta_0^2 = -\frac{1}{2}\omega^2 m_e / E$ are introduced.

The solution of equation (24) satisfies the condition $u(r = r_o) = 0$ at $r_o = \alpha/\beta_0^2 = -Ze^2/E = Ze^2/|E|$, which corresponds to the case when the boundary condition (7.5) holds. Under this condition, the wave amplitude becomes zero at $r = r_o$, where, respectively, the electron trajectory appears (the radius $r_o$ should be determined). Allowing for the asymptotic behavior of the wave at $r \to \infty$, we write the general solution (24) as $u = u_-(r) + u_+(r) = e^{-k_0 r} f_-(r) + e^{k_0 r} f_+(r)$. By substituting it in (24), we obtain the following equations:

$$f_\pm''(r) \pm 2k_0 f_\pm'(r) + \frac{\beta_1}{r_0 - r} f_\pm(r) = 0, \tag{25}$$

where $\beta_1 = k_0^2 \alpha / \beta_0^2 = \frac{1}{2} Ze^2 \omega^2 m_e / E^2$.

The nontrivial solutions of (25) exist when the $f_\pm(r)$ functions can be presented as the following power series $f_\pm(r) = \sum_{m=1}^{\infty} a_m^{(\pm)}(r_0 - r)^m$ (in which in fact the electron trajectory becomes localized on the surface with the radius $r = r_o$). Equation (25) after this substitution has the following form

$$\sum_{m=1}^{\infty} m(m-1) a_m^{(\pm)}(r_0 - r)^{m-2} \mp 2k_0 \sum_{m=1}^{\infty} m a_m^{(\pm)}(r_0 - r)^{m-1} + \beta_1 \sum_{m=1}^{\infty} a_m^{(\pm)}(r_0 - r)^{m-1} = 0,$$

$$\sum_{m=1}^{\infty} [(m+1)m a_{m+1}^{(\pm)} \mp 2k_0 m a_m^{(\pm)} + \beta_1 a_m^{(\pm)}](r_0 - r)^{m-1} = 0, \tag{26}$$

where the coefficients of the $a_{m\geq 1}^{(\pm)}$ series satisfy the recurrence relation

$$(n+1)n a_{n+1}^{(\pm)} \mp 2k_0 n a_n^{(\pm)} + \beta_1 a_n^{(\pm)} = 0. \tag{27}$$

Thus we have

$$a_{n+1}^{(\pm)} = \Lambda_{n+1}^{(\pm)} a_n^{(\pm)}. \tag{28}$$

where $\Lambda_{n+1}^{(\pm)} = \dfrac{\pm 2k_0 n - \beta_1}{(n+1)n}$. Hence we find that at $n = \beta_1/(2k_o)$ the stable (finite) motion of the electron appears leading to the following solution

$$u_{+,n}(r) = C\exp\{k_{o,n}r\}\sum_{m=1}^{n} a_m^{(+)}(r_{o,n} - r)^m, \qquad (29)$$

where $a_m^{(+)} = 0$ at $m \geq n+1$, C - is a constant. The relationship between the frequency and energy (14.1) $\omega^2 = (2E/\hbar)^2$ is also taken into account in the relations $k_0^2 = -\tfrac{1}{2}\omega^2 m_e/E$ and $\beta_1$, we obtain for the radius of the n-th state

$$r_{o,n} = 2\hbar^2 n^2/(Ze^2 m_e), \qquad (30)$$

and taking into account the relation $E^3/\omega^2 = -\tfrac{1}{8}Z^2 e^4 m_e/n^2$, we find the value of the energy of the n-th state

$$E_n = -\dfrac{Z^2 e^4 m_e}{2\hbar^2}\dfrac{1}{n^2}. \qquad (31)$$

Using the properties of Wronskian for the second-order equation we find the independent solution for the coupled wave of the electron

$$u_{-,n}(r) = u_{+,n}(r)\begin{cases}\displaystyle\int_0^r \dfrac{dr'}{u_{+,n}^2(r')}, & (r < r_{+,n}) \\ \displaystyle\int_r^\infty \dfrac{dr'}{u_{+,n}^2(r')}, & (r > r_{+,n}),\end{cases} \qquad (32)$$

decaying exponentially with distance from the nucleus of the atom $u_{-,n}(r \to \infty) \sim \exp\{-k_{o,n}r\}$. Note that the wave $u_{-,n}(r)$ changes its sign upon the transition r through the point $r_{o,n}$. In accordance with condition (7.5), this indicates the presence of the electron trajectory in this point. Note that the energy of the n-th state coincides exactly with the solution obtained in the Bohr model or in the solution of stationary Schrödinger equation. The general solution for electron wave of the n-th state has the following form:

$$V_n(r,t) = u_{-,n}(r)\dfrac{\cos\omega_n t}{r}. \qquad (33)$$

Note that according to Eqs. (30)-(33), Fig.1 shows the solutions for the electron wave for three lowest stationary states (n=1,2,3). Interestingly, starting from the second lowest state the wave amplitude crosses the zero value more than once, however, only at $r = r_{o,n}$ the derivative of the wave $\frac{\partial}{\partial r} V_n(r,t)$ changes the sign of this point. According to (7.3), this indicates the presence of the electron trajectory only on the surface with the radius $r_{o,n}$. The properties of the trajectory and waves $V_n(r,t)$ described above indicate the spatial location of the electron in the atom hydrogen differing from the known picture described by the Schrödinger wave function. Using these results, we limit ourselves to the discussion of several most significant observations and leave the formulation of a series of problems which can have far going effects for the further investigations.

5. **Discussion and conclusion**

Historically, N. Bohr was the first to explain the spectrum of the hydrogen atom [3] on the basis of using the trajectory classical description of the electron motion complemented by the procedure of quantization of the possible electron orbits. However, using the ideas of Louis de Broglie, E. Schrödinger [5] has explained the spectrum of the hydrogen atom within the purely wave approach, rejecting completely the use of the classical electron trajectory in his equation.

In this work [35], we show for the first time that the spectrum of the hydrogen atom can be described on the basis of the approach, in which the electron wave and its trajectory in the atom are described within the unique approach. The electron trajectory and wave are related to each other, this relationship is described in the V-function method on the basis of the local variational principle. In this approach, the behavior of the electron in the n-th stable spherically symmetrical state is described by the wave $V_n(x,t)$ (33). Its amplitude goes through zero on the sphere with the Bohr radius $r_{o,n}$. This means that there is the electron trajectory on the sphere with this radius (spherical asymmetrical states of the electron motion are considered in the following work). It should be noted that in the paper of N. Bohr the electron moved on the orbit with the fixed distance from the nucleus. The result we obtained only partially reproduces the N. Bohr picture of the hydrogen atom since the electron trajectories "are smeared" over the sphere surface with the Bohr radius $r_{o,n}$ so the electron location becomes distributed over the whole sphere surface. Such a behavior of the electron in a sense is analogous to the spatial delocalization of the electron inside the three-dimensional cloud of the Schrödinger wave function, i.e., the appearance of the electron trajectory as a sphere reflects the non-classical behavior of the electron as a particle if one compares its behavior with the classical Rutherford picture for the hydrogen atom. When this work have been finished, we learned results of B.N. Rodimov published in [36] where the author described the spectrum of atom hydrogen by using stationary equation which formally coincides with our Eq. (21). He obtained the equation in different physical approach based on the special relativity theory with additional 4 postulates. Moreover, he attributes a meaning of some additional force to the used electron wave so he didn't found the electron trajectory which is proposed in our work (for example in contrast to our spherical trajectories, he wrote that the electron moves along the radius of hydrogen atom in symmetric states) .

One can say that the picture of the behavior of the electron trajectories in the hydrogen atom predicted in this work has definite aspects of the model of the Bohr atom and Schrödinger wave theory. It should be noted as well that according to the solution (30), the electron motion in the n-th state of the hydrogen atom obtains the purely wave character, since the electron trajectory "is frozen" on the sphere with the fixed radius in the absence of external fluctuations. The predicted spatial structure of the electron trajectories in a somewhat different manner shows the uncertainty in the behavior of the momentum and coordinate of the electron on the surfaces of the sphere of the stationary state if compared with the known uncertainty relation inherent to the electron according to solution of the Schrödinger equation. A more complete understanding of the trajectory and wave aspects of the behavior of the electron, the same as of other particles, requires the performance and studies of new experiments, for example, using the facilities of modern optics. In this respect we consider that equations (7.2) and (19) reflect the wave nature of the electron motion and they, most probably, can bridge many results obtained earlier in quantum mechanics. For example, it should be noted that the general solution of equation (19) for the wave can contain a superposition of the waves of the type (30) with some weights differing in the energy in the general case. This indicates that the superposition principle studied in detail in quantum mechanics holds. The interference of the electron waves oscillating with different frequencies leads to the appearance of the electron trajectories as 2D surfaces oscillating in time around the atomic nuclear. Understanding of the dynamics and spatial form of the non-stationary trajectories—surfaces of the electron requires the studies of the spatial properties of the destructive interference of the electron waves from different states that obviously has a close analogy with the appearance of the non-stationary oscillating dipole moments in the atoms studied in the coherent transition phenomena [37,38].

The principle difference of the picture of the hydrogen atom we considered from the well known result of the Schrödinger quantum theory is that the predicted electron trajectories in the hydrogen atom are uniformly distributed over the spherical surface with the fixed radius rather than in the 3D cloud of the Schrödinger wave function. We consider that to test the proposed theory one should first of all check the predicted distribution of the electron trajectories on the spherical surfaces related to the stationary quantum states of the hydrogen-like atom. At present, this experiment can be implemented using the possibilities of the modern experimental physics, in particular, scanning tunneling and force microscopy [39, 40], which allows the detailed analysis of the spatial features of the electron motion in the atom. However one should take into account that the influence of additional interactions which will make an influence on the spatial structure of the electron trajectory surfaces that is a subject of special investigation.

In summary, note that in this work we used the approach to the cognition of the nature of the electron and its manifestations not on the basis of the possibilities of the present measurement methods but accepting its unique physical nature, which contains its wave essence and corpuscular (trajectory) way being without contradictions. The relations linking the properties of the wave and particle (14.1), (14.2), as well as the new wave equation (19) and its solutions (31), (32) reveal the ontological contents of the developed theory, having the direct relationship to the new continuation of the optical-mechanical analogy deprived of the dualistic approach to the description of the corpuscular and wave properties of the electron. We consider that the theory proposed in this work allows one to make a new experimental test of the standard quantum theory and to shed new light on the fundamental nature of an electron.

Captions for the figures.

Figure 1. Electron wave amplitude $R_1$ (for principle number n=1) as a function of relative distance $r/r_o$, here $r_o$ is a Bohr radius of 1th stationary state.

Figure 2. Electron wave amplitude $R_2$ (for principle number n=2) as a function of relative distance $r/r_o$, here $r_o$ is a Bohr radius of 2th stationary state.

Figure 3. Electron wave amplitude $R_3$ (for principle number n=3) as a function of relative distance $r/r_o$, here $r_o$ is a Bohr radius of 3th stationary state.

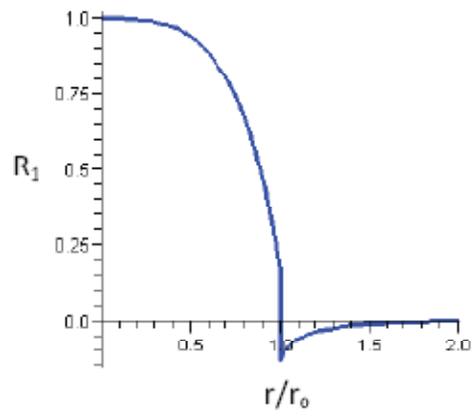

Figure 1.

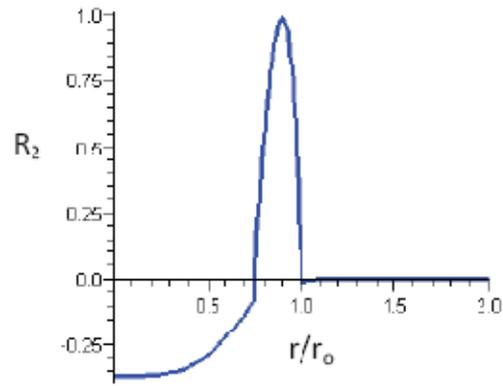

Figure 2.

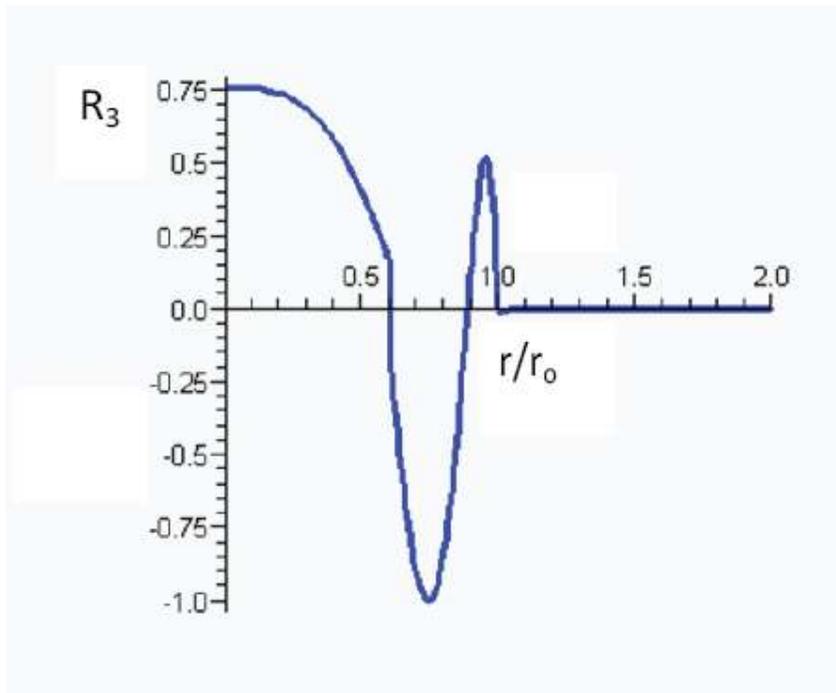

Figure 3.